\documentclass[aps,prb,twocolumn,superscriptaddress,shopacs]{revtex4}

\usepackage{graphicx,color,ulem}
\begin{document}


\title{Magnons and continua in a magnetized and dimerized spin-1/2 chain}

\author{M. B. Stone}
\email{stonemb@ornl.gov}
\affiliation{Quantum Condensed Matter Science Division, Oak Ridge National
Laboratory, Oak Ridge, Tennessee 37831, USA}
\affiliation{Department of Physics and Astronomy, The Johns Hopkins
University, Baltimore, MD 21218}

\author{Y. Chen}
\affiliation{National Institute of Standards and Technology,
 Gaithersburg, MD 20899}
\affiliation{Department of Physics and Astronomy, The Johns Hopkins
University, Baltimore, MD 21218}

\author{D. H. Reich}
\affiliation{Department of Physics and Astronomy, The Johns Hopkins
University, Baltimore, MD 21218}

\author{C. Broholm}
\affiliation{Department of Physics and Astronomy, The Johns Hopkins
University, Baltimore, MD 21218}
\affiliation{Quantum Condensed Matter Science Division, Oak Ridge National
Laboratory, Oak Ridge, Tennessee 37831, USA}
\affiliation{National Institute of Standards and Technology,
 Gaithersburg, MD 20899}

\author{G. Xu}
\affiliation{Condensed Matter Physics and Materials Science
Department, Brookhaven National Laboratory, Upton, NY 11973}
\affiliation{Department of Physics and Astronomy, The Johns Hopkins
University, Baltimore, MD 21218}

\author{J. R. D. Copley}
\author{J. C. Cook}
\affiliation{National Institute of Standards and Technology, Gaithersburg, MD
20899}

\date{\today}

\begin{abstract}
We examine the magnetic field dependent excitations of the dimerized spin-1/2 chain, copper nitrate, with antiferromagnetic intra-dimer exchange $J_1=0.44$ meV and exchange alternation $\alpha=J_2/J_1=0.24$. Magnetic excitations in three distinct regimes of magnetization are probed through inelastic neutron scattering at low temperatures.  At low and high fields there are three and two long-lived magnon-like modes, respectively. The number of modes and the anti-phase relationship between the wave-vector dependent energy and intensity of magnon scattering reflect the distinct ground states: A singlet ground state at low fields $\mu_0H < \mu_0H_{c1} = 2.8$~T and an $S_z=1/2$ product state at high fields $\mu_0H > \mu_0H_{c2} = 4.2$~T.  In the intermediate field regime, a continuum of scattering for $\hbar\omega\approx J_1$ is indicative of a strongly correlated gapless quantum state without coherent magnons.
\end{abstract}

\maketitle


\section{Introduction}
Insulating solids containing dimers of magnetic atoms with antiferromagnetic interactions can form a non-magnetic singlet ground state. As opposed to conventional dielectrics, the incipient magnetism of inter-atomic singlets can be exposed by chemical doping, pressure, and magnetic fields and is accessible to inelastic thermal neutron scattering. There are thus excellent opportunities to explore quantum collective phenomena at the onset of magnetism in such materials.

A case in point are materials containing dimerized antiferromagnetic spin-1/2 chains.\cite{giamarchi08,sachdevscience2000,nohadani2005,elstner1998,kotov1998} When antiferromagnetic Heisenberg interactions prevail ($J_1>>J_2\equiv \alpha J_1$), these spin system can be driven through {\it two} quantum critical points, both associated with Bose-Einstein condensation of magnons. \cite{ricescience,Matsubara56,affleck1990,Nikuni00,giamarchi1999,Orignac2007,sebastianNature,suh2011,conner2010,zapf2014} Immediately above the lower critical field, $\mu_0H_{c1} \approx J_1(1-\alpha)/(g\mu_B)$, a macroscopic density of magnons gives rise to magnetization.\cite{barnesprb1999,chitraprb1997,yuprb2000,hagaprb2001} The reverse situation pertains at the upper critical field where magnon condensation {\it below} $\mu_0H_{c2}=J_1(1+\alpha)/(g\mu_B)$ reduces the magnetic moment per spin below saturation ($M<M_{\rm sat}\equiv g\mu_B/2$).

While the intermediate field state for the idealized alternating spin chain is predicted to be a Luttinger liquid with dynamic transverse staggered spin correlations, the quantum critical nature of this phase implies sensitivity to sub-leading interactions and anisotropies. These ultimately control the low temperature behavior of real materials  for $H_{c1}<H<H_{c2}$ where N\'{e}el or spin-Peierls order is expected at sufficiently low temperatures.

A number of spin-1/2 dimer compounds have been subjected to detailed thermo-magnetic measurements exploring the full field range.\cite{kageyama1999etal,jaime2004,sebastianNature,tanaka2001,chaboussant1998,nawa2011} To understand spin correlations at the atomic scale, however, requires inelastic neutron scattering experiments, which so far have been limited to the vicinity of the lower critical field $H_{c1}$. \cite{fujita95,leuenbergerprb1985,cavadini99,cavadini2002,ruegg2003} By selecting a material, copper nitrate, with a suitable energy scale and utilizing high efficiency cold-neutron instrumentation, we can provide a comprehensive set of data for the field and and wave vector dependence of collective magnetic excitations in a system of interacting spin-1/2 dimers throughout the relevant field range. While the resolution of the present experiment is insufficient to separate a presumed low energy continuum from intense elastic nuclear scattering, the experiment provides evidence for an extended gapless phase for $H_{c1}<H<H_{c2}$ in the form of a continuum of scattering for $J_1(1-\alpha)<\hbar\omega<J_1(1+\alpha)$.

The paper is organized as follows. Section~\ref{cnintro} provides background information regarding copper nitrate. Section~\ref{exptechniques} describes sample preparation and neutron instrumentation. The experimental data from complementary triple axis and time of flight neutron spectrometers are presented in section~\ref{expresults}. Focusing in succession on each regime of the phase diagram, section~\ref{discussion} discusses the data. An exact diagonalization study, which is a useful point of reference opens this section. A summary of the conclusions from this work is presented in section~\ref{conclusions}.

\section{Copper Nitrate}
\label{cnintro}

\subsection{Alternating spin chain model}

The strongest interactions linking spin-dimers in copper nitrate (Cu(NO$_{3}$)$_{2}\cdot$2.5D$_{2}$O, (CN)) create a quasi-one-dimensional spin system that can be described as an alternating 1d antiferromagnetic chain. In an external magnetic field ${\mathbf H}=H\hat{z}$, the spin-Hamiltonian is
\begin{equation}
{\mathcal H} = \sum_{n} J_{1}({\mathbf S}_{2n}\cdot {\mathbf S}_{2n+1} +\alpha {\mathbf S}_{2n+1}\cdot {\mathbf S}_{2n+2}) - g\mu_{B}\mu_0{\mathbf H}\cdot {\mathbf S}_{n},
\label{eq:altchainh}
\end{equation}
where $J_{1}$ is the intra-dimer exchange, $\alpha$ is the ratio of inter- and intra-dimer exchange $\alpha = J_{2}/J_{1}$, $g$ is the Land$\mathrm{\acute{e}}$ g-factor,  $\mu_B$ is the Bohr magneton, and $\mu_0$ is the vacuum permeability.  Other examples of antiferromagnetic alternating chains include (VO)$_{2}$P$_2$O$_{7}$, \cite{johnston87,eccleston94,garrett97prb,garrett97prl} CuWO$_{4}$,\cite{lake96} and Cu(L-aspartato)(H$_{2}$O)$_{2}$.\cite{calvo99}

Early thermo-magnetic measurements on CN  showed exponentially activated ($\exp(-\Delta/T)$) susceptibility and specific heat data  indicative of an energy gap, $\Delta$, in the magnetic excitation spectrum.\cite{bergerpr1963,friedbergjappl1968,myers1969} These results and nuclear magnetic resonance (NMR) measurements \cite{wittekoek1968} were analyzed in terms of isolated dimers with weak inter-dimer interactions.  The first accounts of inter-dimer interactions creating a 1d system resulted from an analysis of the crystal structure,\cite{morosin70} and additional thermo-magnetic studies.\cite{bonnerearly}  Two models emerged: an alternating chain model and a two-leg ladder model.  The alternating chains were proposed to extend along the $[101]$ direction, while the crystalline $b$-axis was proposed as the extended direction for the 2-leg ladder model.  The originally proposed alternating chain was mistakenly projected onto the $[101]$ axis.  Both models identify the spin-dimer as composed
of nearest neighbor copper sites with the intra-dimer vector ${\mathbf d}_{1}$ (Fig.~\ref{fig:cnstructfig}(c)). NMR measurements as a function of applied field and orientation showed the inter-dimer exchange primarily lies in the $ac$ plane forming alternating chains rather than spin-ladders extending along $b$.\cite{diederix78}

Specific heat and NMR studies of CN provide evidence of a field induced phase transition to long range antiferromagnetic order for $T\leq 175$~mK in magnetic fields between $\mu_0 H_{c1}=2.8$ T and $\mu_0 H_{c2}=4.2$ T. \cite{vantolphysica1973}   A model to account for these phase transitions was developed by mapping the system
in the intermediate field regime onto a $S=\frac{1}{2}$ linear chain in an effective magnetic field $H_{E}$, such that $H_{E}=0$ when
$H=\frac{1}{2}(H_{c1}+H_{c2})$.\cite{tachikisuppl1970}  Subsequent thermo-magnetic measurements were interpreted with moderate success based on this approach.\cite{diederix78,diederixphys77}   An elastic neutron scattering experiment investigating the ordered phase of CN supported the alternating chain model and confirmed the ordered magnetic structure proposed by Diederix \textit{et al}.\cite{eckert79} However, as we shall see in
section~\ref{detailed_structure_section}, Diederix's alternating chain model incorrectly placed the chain axis along the $[101]$ direction.

The aforementioned measurements led to general acceptance of the alternating chain model for CN, and established the exchange parameters to be given by $J_{1}\approx$ 0.44 meV and $\alpha \approx$ 0.24.  A direct spectroscopic measurement of the magnetic excitation spectrum was subsequently reported by Xu \textit{et al}.\cite{gxu00}  These measurements showed the dynamic spin correlation function ${\mathcal S}(\tilde{q}, \hbar\omega)$ of the singlet-triplet excitations is well described by the single mode approximation (SMA) with dispersion relation
\begin{equation}
\epsilon({\mathbf Q}) = J_{1}-\sum_{\bf u}\frac{J_{\bf u}}{2}\cos({\mathbf Q}\cdot{\bf u}).
\label{eq:altchaindisp}
\end{equation}
Here $J_{1}$ is the intra-dimer exchange and $J_{\bf u}$ the interdimer exchange.\cite{gxu00}  Wave-vector dependent two-magnon excitations were also observed at higher energy transfers using inelastic neutron scattering.\cite{tennantcontinuum}  More recently, the zero-field single and multimagnon interactions have been examined with high resolution inelastic neutron scattering techniques to illustrate the influence of temperature on the quasiparticle excitations.\cite{tennantspinecho}

\subsection{Crystal Structure}
\label{detailed_structure_section}
 \begin{figure*}
 \centering\includegraphics[scale=1]{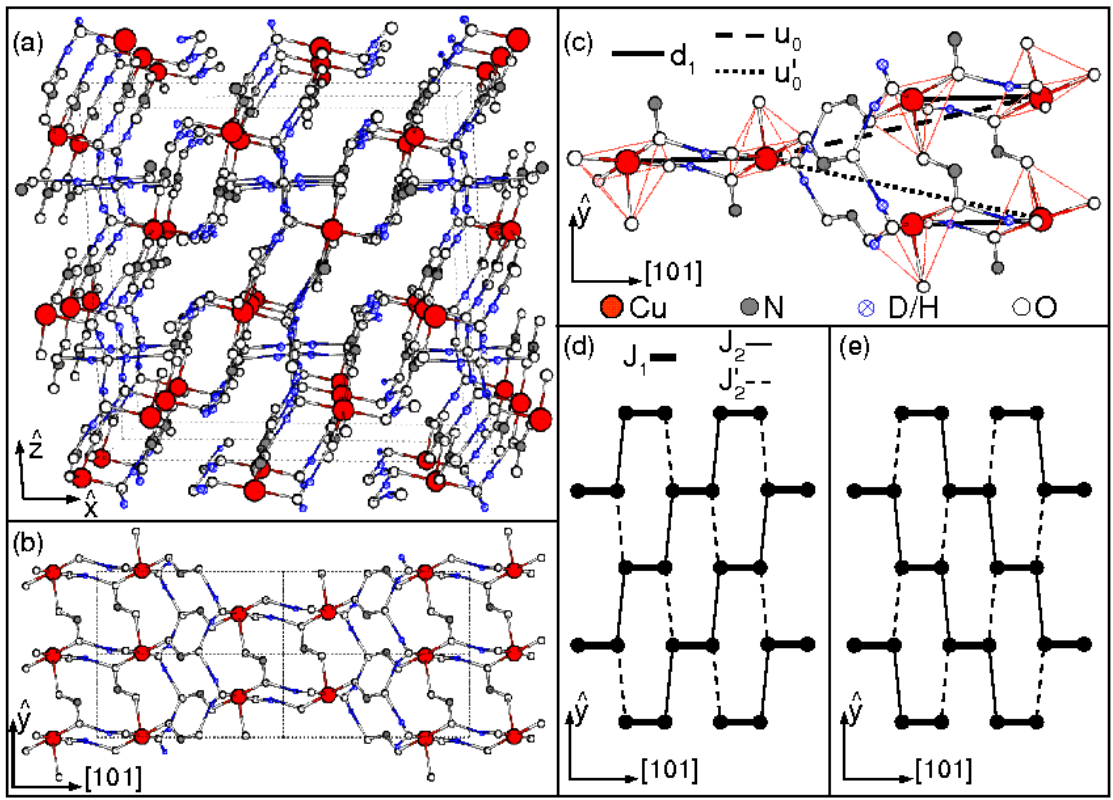}
 \caption{\label{fig:cnstructfig}
(color online) Crystal structure of CN.  (a) Perspective view along the $b$-axis.  (b)  View along the $[10\bar{1}]$ direction.  (c) Expanded perspective view along the $[10\bar{1}]$ direction with 4+1 coordination of Cu$^{2+}$ sites highlighted by square pyramidal polyhdedra.  Vectors $\mathbf{d}_1$, $\mathbf{u}_0$ and $\mathbf{u}_0^{\prime}$ are the respective intra- and two possible inter-dimer vectors described in the text.  The interdimer exchange path we propose connects dimer units for which the basal planes of the 4+1 coordination are nearly co-planer along the [111] direction in (b), (c) and (d) via the $\mathbf{u}_0$ interdimer bond and illustrated by $J_1$-$J_2$ alternating chains in (d) and (e). (d) and (e) Schematic of interdimer interactions in CN
in adjacent $[10\bar{1}]$ planes. intra-dimer interactions, $J_1$, are represented by circles linked by solid heavy lines.  The two possible interdimer interactions are represented as solid ($J_2$) and dashed ($J_2^{\prime}$) lines.  In (b) and (c), proton sites D/H(4) and D/H(5), and oxygen sites O(3), O(4), and O(9) as enumerated in Ref.~\onlinecite{morosin70} have been omitted for clarity.  These omitted atoms participate in bonding with water sites adjoining the copper atom, secondary exchange paths along the $b$ direction, and bonding between the $[10\bar{1}]$ planes.}
\end{figure*}

CN is monoclinic with space group $I2/c$ \cite{spacegroupnote} and room temperature lattice constants $a=16.45$~\AA, $b=4.94$~\AA, $c=15.96$~\AA, and $\beta=93.77^{\circ}$.\cite{morosin70,garaj1968,raj1982} We find the low-temperature, $T\leq 4$~K, lattice constants to be $a=16.12$~\AA, $b=4.90$~\AA, $c=15.84$~\AA, and $\beta=92.90^{\circ}$.  Depicted in Fig.~\ref{fig:cnstructfig}, the crystal structure features Cu$^{2+}$ ions in a distorted 4+1 tetrahedral coordination with five oxygen atoms, (Fig.~\ref{fig:cnstructfig}(c)).  The four basal plane Cu-O bonds each have a bond length of approximately 2~\AA\ while the apical Cu-O bond length is 2.39~\AA.

The $J_1$-coupled dimer is formed by neighboring Cu$^{2+}$ ions where adjacent pairs of the short bond coordinated oxygen atoms form almost parallel planes that establish two Cu-O-H-O-Cu linkages as shown in Fig.~\ref{fig:cnstructfig}(c).  This exchange path is hydrogen bond mediated via a water group coordinated to one copper atom and a nitrate group coordinated to the other copper atom.

The alternating chain model involves structures extending along the [111] and [1$\bar{1}$1] directions with Cu-O-N-O-H-O-Cu interdimer contacts.  As shown in Fig.~\ref{fig:cnstructfig}(b) and schematically in Fig.~\ref{fig:cnstructfig}(d)-(e), each $[10\bar{1}]$ plane contains  interpenetrating [111] and [1$\bar{1}$1] alternating chains.  Though the sequence of atomic contacts are identical for both chains in a single $[10\bar{1}]$  plane, the bond angles and distances are different.  We denote the two potential inter-dimer exchange constants as $J_2$ and $J_2^{\prime}$.  Identical interactions for the two interpenetrating
chains ($J_2 = J_2^{\prime}$) would produce a two-dimensional distorted square lattice rather than alternating chains.

In the alternating chain model, $J_2 \gg J_2^{\prime}$, such that only one type of magnetic chain, [111] or [1$\bar{1}$1], populates a single $[10\bar{1}]$ plane, while the other type of chain populates neighboring planes as shown in Figs.~\ref{fig:cnstructfig}(d)-(e).

The fractional coordinates of the intra-dimer vector is ${\bf d}_{1}$ = [0.25, $\pm$0.03, 0.23] ($|{\bf d}_{1}|$=5.33 \AA), and the center to center inter-dimer vectors are ${\bf u}_0$, ${\bf u}_0^{\prime} = [1, \pm 1, 1]/2$ ($|{\bf u}_{0}|= |{\bf u}_0^{\prime}| =11.351$ \AA).  The $[111]$ alternating chain in Fig.~\ref{fig:cnstructfig}(c) (dashed line) connects dimer units with neighboring 4+1 oxygen coordinated basal planes that are nearly co-planer while the $[1\bar{1}1]$ alternating chain in Fig.~\ref{fig:cnstructfig}(c) (dotted line) connects dimer units with neighboring 4+1 oxygen coordinated basal planes that are nearly parallel but \textit{not} co-planer.

Based on experience with other Cu$^{2+}$ systems with similar coordination, and the observation that the dominant $3d$ electron density should reside in the basal plane for this coordination, we hypothesize that the dashed exchange path, ${\bf u}_0$, in Fig.~\ref{fig:cnstructfig}(c), mediates the dominant inter-dimer exchange interaction.  This corresponds to a nearest inter-dimer spin-spin distance of 6.22~\AA, as opposed to 6.32~\AA\ for the other exchange path. However, because the neighboring $[10\bar{1}]$ planes contain identically coordinated alternating chains in the other direction, neutron scattering can not distinguish
between these two types of coordination. As discussed below, we are however, able to place limits on the amount of two-dimensional interchain exchange, $J_2^{\prime}$.

\section{Experimental techniques}
\label{exptechniques}

\subsection{Sample preparation}
Deuterated CN single crystals were grown by cooling a saturated solution of deuterated CN in D$_{2}$O.\cite{gxuthesis} Deuterated CN was obtained through repeated distillation (3 - 4 times) of commercially available hydrogenous CN in high-purity D$_{2}$O ($> 99.9 \%$).  Seed crystals were initially grown by rapid cooling of  saturated solutions from $80 ^{\circ}$C to $30 ^{\circ}$C.  Single crystals were subsequently grown from suspended seeds in filtered saturated solutions of CN by slowly cooling from $50 ^{\circ}$C to $30 ^{\circ}$C.  The sample used for inelastic neutron scattering consisted of three 94$\%$ deuterated single crystals with a total mass of 9.25 grams co-aligned to within 1.25$^{\circ}$.

\subsection{Neutron Instrumentation}
Field dependent measurements  were made using the SPINS triple-axis and the DCS time-of-flight spectrometers at NIST. The  $b$-axis was vertical so that ~$\mathbf{Q}$ was in the $(h0l)$ plane with equal contributions to the magnetic scattering from alternating chains extending along the [111] and [1$\bar{1}$1] directions.   The sample environment for both measurements consisted of a dilution refrigerator within a split-coil 11.5 Tesla vertical field cryomagnet.  Scattering intensities were normalized to the incoherent elastic scattering from a standard vanadium sample to provide a measurement in absolute units of the following two-point dynamic spin correlation function:
\begin{eqnarray}
{\cal S}^{\alpha\beta}({\bf Q}, \omega)&=&
\frac{1}{2\pi\hbar}\int_{-\infty}^\infty dt\ e^{-i\omega t}
\frac{1}{N}\sum_{{\bf r}{\bf r}^\prime}e^{i{\bf Q}\cdot({\bf r-r^{\prime}})}\nonumber\\
&&\langle S^\alpha_{\bf r}(t)S^\beta_{{\bf r}^\prime}(0)\rangle
\end{eqnarray}
Here $\hbar\omega$ and $\hbar{\bf Q}$   are energy and momentum transfer to the sample, $\alpha$ and $\beta$ indicate cartesian components, $N$ is the number of Cu$^{2+}$ sites in the sample, and $S_{\bf r}^{\alpha}$ indicates the $\alpha$ component of the spin operator at position ${\bf r}$. In the following we shall use  $\tilde{q} = {\bf Q} \cdot {\bf u}_0$ to indicate wave-vector transfer along the chains.

The normalized magnetic portion $\tilde{I}({\bf Q},\hbar\omega)$ of the inelastic neutron scattering intensity  is  related to the dynamic spin correlation function as follows
\begin{eqnarray}
\tilde{I}({\bf Q},\hbar\omega) = \int d^3{\bf Q}^{\prime} \hbar d\omega^{\prime}{\mathcal {R}}_{{\mathbf{Q}},\omega} (\mathbf{Q}-\mathbf{Q}^{\prime},\omega-\omega^{\prime}) \nonumber \\
|\frac{g}{2}F(\mathbf{Q}^{\prime})|^2 \sum_{\alpha\beta}(\delta_{\alpha\beta}-\hat{Q^{\prime}}_\alpha\hat{Q^{\prime}}_\beta){\mathcal {S}^{\alpha\beta}}(\mathbf{Q}^{\prime},\omega^{\prime}),
\label{eq:intneut}
\end{eqnarray}
where $F({\bf Q})$ is the magnetic form factor of the Cu$^{2+}$ ion and ${\mathcal {R}}_{{\mathbf{Q}},\omega}$ is the  instrumental resolution function satisfying the following normalization condition.\cite{chesseraxe}
\begin{equation}
 \int d^3{\bf Q}^{\prime} \hbar d\omega^{\prime}{\mathcal {R}}_{{\mathbf{Q}},\omega} (\mathbf{Q}-\mathbf{Q}^{\prime},\omega-\omega^{\prime}) =1
\end{equation}

    \begin{table*} \centering
    \begin{tabular}{|c|r@{.}l|r@{.}l|c|c|c|c|c|c|}\hline
        \multicolumn{1}{|c|}{ Configuration}     &
        \multicolumn{2}{c|}{ E$_{i}$ (meV)}     &
        \multicolumn{2}{c|}{ \textit{f} (Hz)}   &
        \multicolumn{1}{c|}{ resolution mode}  &
        \multicolumn{1}{c|}{ $\hbar\omega_{\mathrm{min}}$ (meV)}     &
        \multicolumn{1}{c|}{ $\hbar\omega_{\mathrm{max}}$ (meV)}     &
        \multicolumn{1}{c|}{ FWHM$\delta\hbar\omega$ ($\mu$eV)}   &
        \multicolumn{1}{c|}{ $|\mathbf{Q}_{\mathrm{min}}|$ (\AA$^{-1}$)}    &
        \multicolumn{1}{c|}{ $|\mathbf{Q}_{\mathrm{max}}|$ (\AA$^{-1}$)}
        \\ \hline

 1 & 2&10  & 333&3  & medium & -0.12  & 1.10  & 28 & 0.18 & 1.79 \\
 2 & 1&90  & 333&3  & low    & -0.15  & 0.95  & 50 & 0.17 & 1.70 \\
 3 & 1&132 & 266&8  & low    & -0.04  & 0.57  & 29 & 0.13 & 1.31 \\
 4 & 1&90  & 279&95 & low    & -0.15  & 1.06  & 57 & 0.17 & 1.70 \\
 \hline
 \end{tabular}
 \caption{Experimental configurations for measurements using the DCS at NIST. The frequency of neutron pulses at the sample is $f$. The resolution mode is determined by the widths of the chopper slots phased to transmit the beam; see Ref.~\onlinecite{copleyphysica1992}.  Minimum and maximum energy transfers correspond to first and last channels in the time-of-flight histogram. The FWHM $\delta\hbar\omega$ is the energy resolution at the elastic position. The minimum and maximum wave vector transfers are also given. The configuration number is referenced in figure captions.}
 \label{tab:dcsconfigs}
 \end{table*}

\subsubsection{Triple-Axis Configurations}
To probe inter-chain interactions we used a conventional cold neutron triple axis configuration on the SPINS instrument at NIST with fixed final energy $E_{f}$=2.5 meV. A cooled Be filter was placed after the monochromator, and the horizontal collimation was $50^{\prime}/k_{i}(\AA^{-1})$-80$^{\prime}$-40$^{\prime}$-80$^{\prime}$, where $k_{i}$ is the incident wave-vector.  This configuration resulted in a measured full width at half maximum (FWHM) elastic energy resolution of $\delta \hbar\omega=50$~$\mu$eV.

For high efficiency mapping of ${\cal S}({\bf Q} \omega)$, a 256 channel 25 cm wide by 20 cm tall position sensitive detector (PSD) was placed after a 23 cm wide by 15 cm tall flat pyrolytic graphite analyzer (PG(002)).  Setting the final energy $E_{f}=2.8$~meV for the centerline of the PSD resulted in sensitivity to a range of final energies $2.4$~meV~$< E_{f} < 3.3$~meV, determined by the range of scattering angles subtended by the analyzer and detector.

The horizontal beam divergence around the monochromator was $50^{\prime}/k_{i}(\AA^{-1})$-83$^{\prime}$ and a cooled Be filter was placed between the monochromator and sample.  An $80^{\prime}$ horizontal radial collimator with a focal length matching the beam line distance to the sample was between the analyzer and PSD covering their full angular acceptance.  For a given setting of the sample and instrument, neutrons with a range of energies and momentum transfer were detected by the PSD.  The corresponding deviation in $\mathbf{Q}$ was parallel to the face of the analyzer.\cite{shleeprl2000} With the (101) direction normal to the analyzer face, the projection of wave vector transfer along the chains, $\tilde{q}$, was constant across the PSD.

With this configuration a region of $\tilde{q}-\hbar\omega$ space was efficiently mapped out by fixing $E_i$ and $E_f$ and scanning the central scattering angle, $2\theta$, while maintaining the (101) direction perpendicular to the analyzer by a corresponding rotation of the sample.  $2 \theta$ scans in the range of 10$^{\circ}$ to 66$^{\circ}$ were performed with incident energies of $E_{i}=$2.83, 3.05, 3.25, and 3.55 meV.  This resulted in an average FWHM energy resolution of $\delta E$ = 0.12 meV and FWHM  $\mathbf{Q}$ resolution along the projected chain direction of $\delta Q_{\parallel}$ = 0.03 \AA$^{-1}$.\cite{chesseraxe}

\subsubsection{Time-of-flight Configurations}
The DCS has three banks of detectors, one below, one above, and one centered in the horizontal plane, with a total of 913 individual detectors located tangent to Debye-Scherrer cones.\cite{copleyphysica1992}  The $[101]$ sample axis was oriented along the incident beam direction. All data correspond to wave vector transfers along the $(101)$ projected chain axis. Incident wavelengths, chopper frequencies, and resolution modes were chosen in consideration of the resulting intensity at the sample, energy resolution, and ranges of energy transfer and wave vector transfer. The four instrument configurations listed in Table~\ref{tab:dcsconfigs} were employed. An oscillating $2^{\circ}$ radial collimator was in place between the sample and detectors to absorb neutrons scattered from material outside the sample volume.

\section{Experimental Results}
\label{expresults}

\subsection{Interchain coupling}
\label{interchain}
To determine the spin Hamiltonian for CN a zero field inelastic neutron scattering experiment was conducted on the SPINS triple-axis spectrometer at NIST with the sample oriented in the $(hkh)$ zone.
\begin{figure}
\centering\includegraphics[scale=0.75]{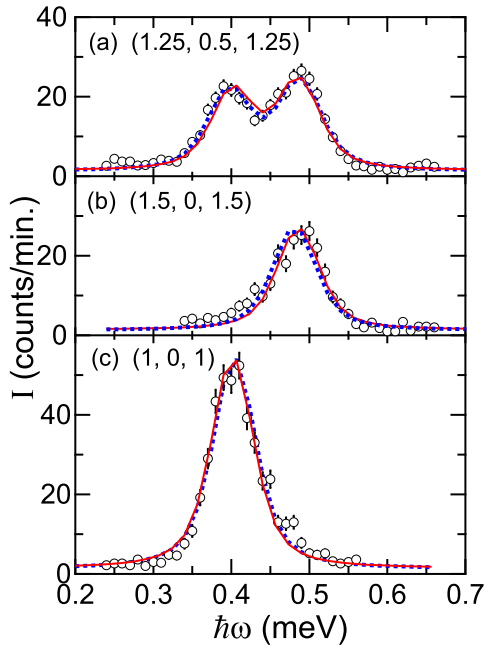}
\caption{\label{fig:1dor2dplot}
(color online) Constant $\mathbf{Q}$ scans for CN at $T=1.25$ K.  Red solid lines are from a global fit to resolution convolved single modes with dispersion as listed in Eq.~\ref{eq:altchaindisp}.  Blue dashed lines are from a global fit to resolution convolved single modes with dispersion as listed in Eq.~\ref{eq:twoddisp}.}
\end{figure}
Figure~\ref{fig:1dor2dplot} shows three constant $\mathbf{Q}$ scans for $T=1.25$~K, which is well above the critical temperature for magnetic ordering. Two modes are observed at (1.25, 0.5, 1.25) (Fig.~\ref{fig:1dor2dplot}(a)), while only a single mode is observed at (1.5, 0, 1.5) and (1, 0, 1), (Figs.~\ref{fig:1dor2dplot}(b) and (c)).  The peaks in Figs.~\ref{fig:1dor2dplot}(b) and (c) are at similar energy transfers as the peaks in Fig.~\ref{fig:1dor2dplot}(a).  Furthermore, all of the peaks in Fig.~\ref{fig:1dor2dplot} are located at the top or the bottom of the band of magnetic excitations previously measured for CN.\cite{gxu00}  Based on a Gaussian peak approximation, the mode energies for the data in Fig.~\ref{fig:1dor2dplot} are $\hbar \omega$ = 0.394(2)meV and 0.487(2) meV, and the energies of the peaks in Figs.~\ref{fig:1dor2dplot}(b) and
(c) are $\hbar \omega$ = 0.488(2) meV and 0.402(1) meV respectively.

For the [1 $\pm$1 1] alternating chain model, there are two distinct modes propagating in the $(hkh)$ plane.
Here denoted A and B, these correspond to the two directions of alternating chains, [111] and [1$\bar{1}$1]. Considering interactions, $J_2^{\prime}$, between [1 $\pm$1 1] chains gives the following dispersion relations.
 \begin{eqnarray}
\epsilon_{{\mathrm A}}(\mathbf{Q}) &=& J_{1} - \frac{J_{2}}{2}\cos(\pi(h+k+l)) \nonumber \\
 & & - \frac{J_{2}^{\prime}}{2}\cos(\pi (h - k + l))\nonumber \\
\epsilon_{{\mathrm B}}(\mathbf{Q}) &=& J_{1} - \frac{J_{2}}{2}\cos(\pi (h - k+ l)) \nonumber \\
  & &  - \frac{J_{2}^{\prime}}{2}\cos(\pi (h + k + l)).
 \label{eq:twoddisp}
 \end{eqnarray}
Note that if the primary chain direction is along [111] ([1$\bar{1}$1]), then interchain interactions are along the [1$\bar{1}$1] ([111]) direction as shown in Figs.~\ref{fig:cnstructfig}(d) and (e).  The difference in mode energies for the wave-vectors in Figs.~\ref{fig:1dor2dplot}(b) and (c) is $J_{2}+J_{2}^{\prime}$, and the difference in energies of the excitations in Fig.~\ref{fig:1dor2dplot}(a) is $J_{2}-J_{2}^{\prime}$.  The measured difference in the energies of the modes in Figs.~\ref{fig:1dor2dplot}(b) and (c) and the modes in Fig.~\ref{fig:1dor2dplot}(a) are 0.086(1) and 0.093(3) meV respectively.  From Eq.~\ref{eq:twoddisp}, the difference in these values limits the magnitude of inter-dimer interactions perpendicular to the primary chain direction to $2J_{2}^{\prime} = -7$(3)~$\mu$eV.  The mean difference in peak locations in Fig.~\ref{fig:1dor2dplot} corresponds to inter-dimer exchange $J_{2}$=0.090(2) meV.

We also performed global fits to these data using the dispersion relations in Eq.~\ref{eq:twoddisp} and a dimer structure factor convolved with the instrumental resolution function (dashed lines in Fig.~\ref{fig:1dor2dplot}).  The results are consistent with the simplified fitting analysis: $J_2^{\prime} = -6(1)$~$\mu$eV, $J_1 = 0.4424(6)$~meV and $J_2 = 0.084(1)$~meV.  A similar global fit with $J_{2}^{\prime}\equiv 0$ and a dimer structure factor yields $J_1 =  0.4436(6)$ meV and $J_2 = 0.085(1)$ meV, solid lines in Fig.~\ref{fig:1dor2dplot}. The values for $J_1$ and $J_2$ are consistent with previous inelastic neutron scattering measurements taking into account the temperature dependent bandwidth normalization factor.\cite{gxu00} Considering the ratio of the weighted means of the exchange parameters determined from the Gaussian peak and resolution convolved fits, $|\bar{J_2}^{\prime}/\bar{J_2}| = 0.06(1)$, it is appropriate to describe CN as well isolated alternating chains that run in the [1~1~1] and [1~$\bar{1}$~1]  directions on adjacent crystal planes spanned by $\bf b$ and $\bf a+b$.  Our analysis of the field dependent excitation spectrum in CN only includes the  $J_1$ and $J_2$ exchange parameters. Although we make structural arguments above concerning the correct inter-dimer linkage, the current experiment does not identify which of the two types of linkage creates the alternating chains, only that one of them corresponds to the alternating chain and the other is very weak.

\subsection{Field Dependent Magnetic Excitations}
\label{finite_field_experiments}

\subsubsection{Triple-axis measurements}
\begin{figure}
  \centering\includegraphics[scale=0.825]{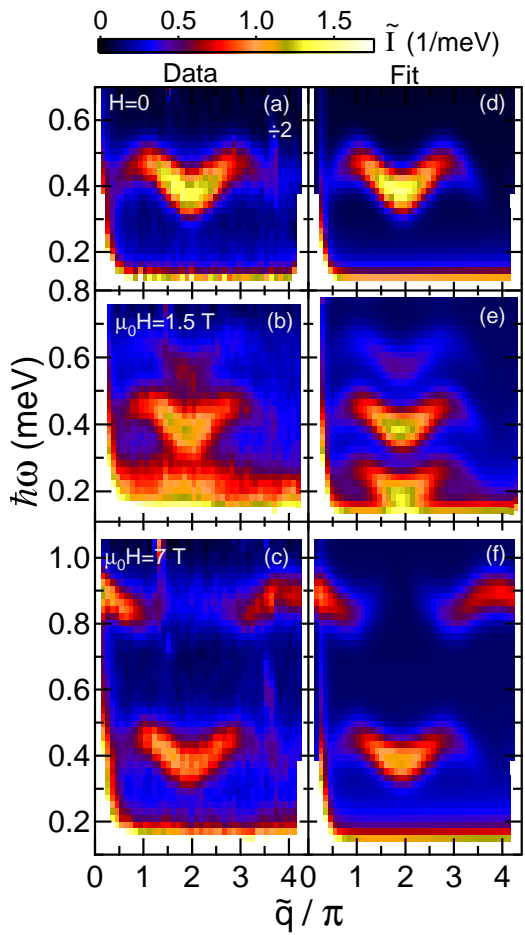}
 \caption{\label{fig:cncontourplotspins}
 (color online) Normalized scattering intensity, $\tilde{I}(\tilde{q}, \hbar\omega)$,  for CN at $T=1.8$ K for (a) $H=0$, (b) $\mu_0 H=1.5$ T, and (c) $\mu_0 H=7$ T from triple-axis measurements.  Zero field results are shown as half the measured/calculated intensity. Panels (d)-(f) are the results of fits described in the text.  }
 \end{figure}

Figures~\ref{fig:cncontourplotspins}(a)-(c) show $\tilde{q}-\hbar\omega$ maps of neutron scattering intensity at $T=1.8$~K measured for $\mu_0 H=0$, $\mu_0 H=1.5$~T, and $\mu_0 H=7$~T on the SPINS triple-axis spectrometer.  In zero magnetic field, there is a single resonant mode with an energy gap of $\Delta\approx 0.35$~meV and a bandwidth of approximately 90~$\mu$eV.  The spectrum appears broadened in comparison to prior zero-field measurements.\cite{gxu00}  This is due to both the elevated temperature and a factor of five coarser energy resolution compared to the prior backscattering measurements.  The periodicity in $\tilde{q}$ of this mode is consistent with Eq.~\ref{eq:twoddisp}.

At $\mu_0 H=1.5$ T, Fig.~\ref{fig:cncontourplotspins}(b) shows three resolution limited modes.  Two of these are clearly seen at $\hbar\omega \approx 0.4$ and $0.6$ meV.  The third is located in the vicinity of $0.2$ meV though it is only apparent through increased scattering intensity in this energy range at field over the corresponding zero-field data. The dispersion of these modes is indistinguishable from the zero field dispersion, $\epsilon_{0}(\mathbf{Q})$, with splitting given by the Zeeman term:
\begin{equation}
\epsilon(\mathbf{Q}) =  \epsilon_{0}(\mathbf{Q}) - g\mu_{B}\mu_0HS_{z}.
\label{eq:zeemandisp}
\end{equation}
For CN in the $(h0l)$ scattering plane with $J_{2}^{\prime}\approx 0$
\begin{equation}
\label{eq:oneddisph0l}
\epsilon_{0}(\mathbf{Q}) =  J_{1} -\frac{J_2}{2}\cos(\pi (h + l)),
\end{equation}
from Eq.~\ref{eq:twoddisp}.  The splitting of the spectrum confirms the singlet-triplet nature of the magnetic excitations in CN as described below in Fig.~\ref{fig:statediagram} for $H<H_{c1}$.

Above the upper critical field, $\mu_0 H_{c2} =4.2$~T, CN is fully magnetized at sufficiently low $T$.  Figure~\ref{fig:cncontourplotspins}(c) shows the $\tilde{q}-\hbar\omega$ dependent scattering intensity for $\mu_0 H=7$~T where two modes are observed.  The lower energy mode has similar intensity modulation and dispersion versus $\tilde{q}$ as in zero field.  In contrast, the $\tilde{q}$ dependence of the upper mode is out of phase with the lower mode.

\begin{figure}
\centering\includegraphics[scale=0.825]{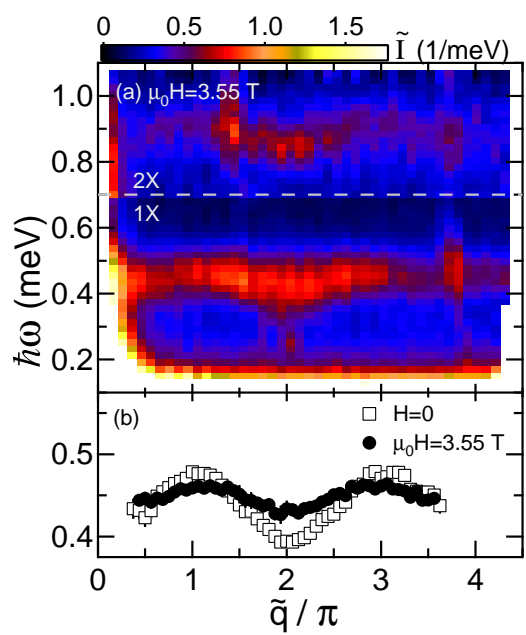}
\caption{\label{fig:cncontour3p55spins}
(color online) (a) Normalized scattering intensity, $\tilde{I}(\tilde{q}, \hbar\omega)$, for CN at $T=1.8$~K and $\mu_0 H=3.55$~T from triple-axis measurements. For clarity, data above the white dashed line ($\hbar\omega > 0.7$~meV) have been multiplied by a factor of two.  (b) Peak positions for $\hbar\omega\approx 0.45$~meV for constant $\tilde{q}$ scans through $H=0$ and $\mu_0 H=3.55$~T spectra at $T=1.8$ K.}
\end{figure}

Data for the intermediate field regime, $\mu_0 H=3.55$~T, is shown in Fig.~\ref{fig:cncontour3p55spins}(a).  The mode at $\hbar\omega\approx 0.4 $ meV is broadened compared to data collected below the lower critical field.  The upper mode of the triplet, although reduced in intensity, has shifted to higher energies with unchanged dispersion and intensity modulation.  Fig.~\ref{fig:cncontour3p55spins}(b) shows fitted peak positions for the zero field and $\mu_0 H=3.55$~T mode, illustrating the change in dispersion at intermediate fields.

\begin{figure}
   \centering\includegraphics[scale=0.825]{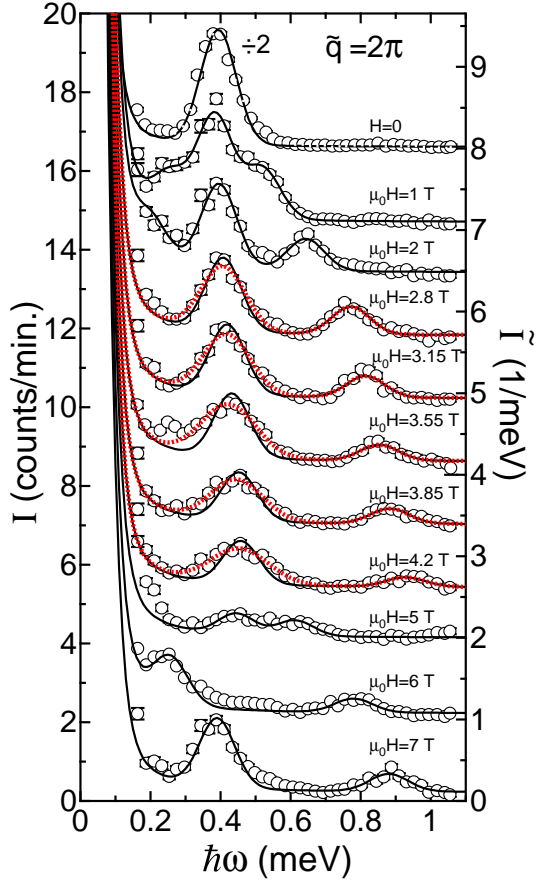}
 \caption{\label{fig:qeq2pi_spins}
(color online) Normalized scattering intensity, $\tilde{I}(\tilde{q}=2\pi, \hbar\omega)$, for CN at $T=1.8$ K for magnetic fields  $0 \leq \mu_0 H \leq 7$ T from triple-axis measurements.  Results have been offset vertically for presentation.  Solid lines are fits to resolution limited Gaussians as described in the text. Dotted red lines for $2.8 \leq \mu_0 H \leq 4.2$ T are fits to resolution limited Gaussians with an additional FWHM added in quadrature as described in the text. Data below $\hbar\omega = 0.15$ meV, which are dominated by incoherent elastic nuclear scattering, have been omitted for clarity.}
\end{figure}

\begin{figure}
\centering\includegraphics[scale=0.825]{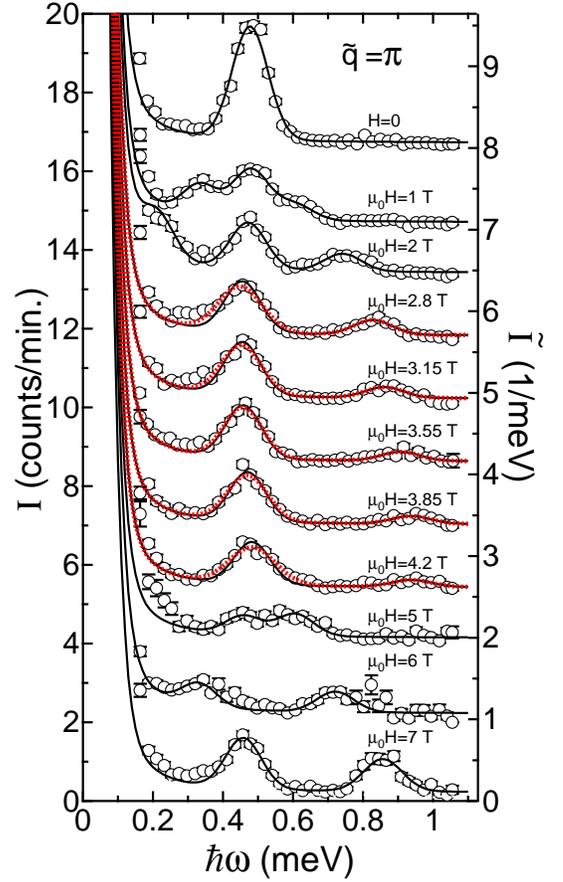}
\caption{\label{fig:qeq2nplus1pi_spins}
(color online) Normalized scattering intensity, $\tilde{I}(\tilde{q}=\pi, \hbar\omega)$, for CN at $T=1.8$ K for magnetic fields  $0 \leq \mu_0 H \leq 7$ T from triple-axis measurements.  Results have been offset vertically for presentation.  Solid lines are fits to resolution limited Gaussians as described in the text.
Dotted red lines for $2.8 \leq \mu_0 H \leq 4.2$ T are fits to resolution limited Gaussians with an additional FWHM added in quadrature as described in the text.
Data below $\hbar\omega = 0.15$ meV, which are dominated by incoherent elastic nuclear scattering, have been omitted for clarity.}
\end{figure}

For a parametric overview of the field dependence, constant $\tilde{q}$ scans were measured at  $\tilde{q}=n\pi$ for integer $n$ for a range of fields covering all portions of the magnetic phase diagram at $T=1.8$ K.  For select values of the applied magnetic field, these are shown in Fig.~\ref{fig:qeq2pi_spins} for $\tilde{q}=2\pi$ and Fig.~\ref{fig:qeq2nplus1pi_spins} for $\tilde{q}=\pi$.

In zero field, there is a single resolution limited mode at the bottom and top of the magnetic band. For magnetic fields up to 2 T, all three of the Zeeman split triplet modes are visible.  However, for larger magnetic fields, the lowest energy excitation of the triplet is obscured by incoherent elastic nuclear scattering.  Above the lower critical field, $\mu_0 H_{c1}$=2.8~T, a mode persists at $\hbar\omega \approx 0.45$ meV, in addition to the higher energy mode which continues to rise in energy. An overall broadening of the $\hbar\omega\approx 0.45$ meV mode is apparent in the intermediate field regime. Comparing the lineshapes for $\tilde{q}=2\pi$ and $\tilde{q}=\pi$,  the $\tilde{q}=2\pi$ scans are consistently broader than the $\tilde{q}=\pi$ scans.  This is also seen in the $\tilde{q}$
dependent spectrum measured at $\mu_0 H=3.55$ T in Fig.~\ref{fig:cncontour3p55spins}(a). At $\mu_0 H=4.2$ T, the peak at $\hbar\omega \approx 0.45$ meV becomes further broadened, and at $\mu_0 H=5$ T a second peak is apparent at $\hbar\omega \approx 0.6$ meV. At $\mu_0 H=6$ and $\mu_0 H=7$ T, the two modes in Figs.~\ref{fig:qeq2pi_spins} and~\ref{fig:qeq2nplus1pi_spins} correspond to those depicted in Fig.~\ref{fig:cncontourplotspins}(c).

\subsubsection{Time-of-flight measurements}
\begin{figure*}
 \centering\includegraphics[scale=0.825]{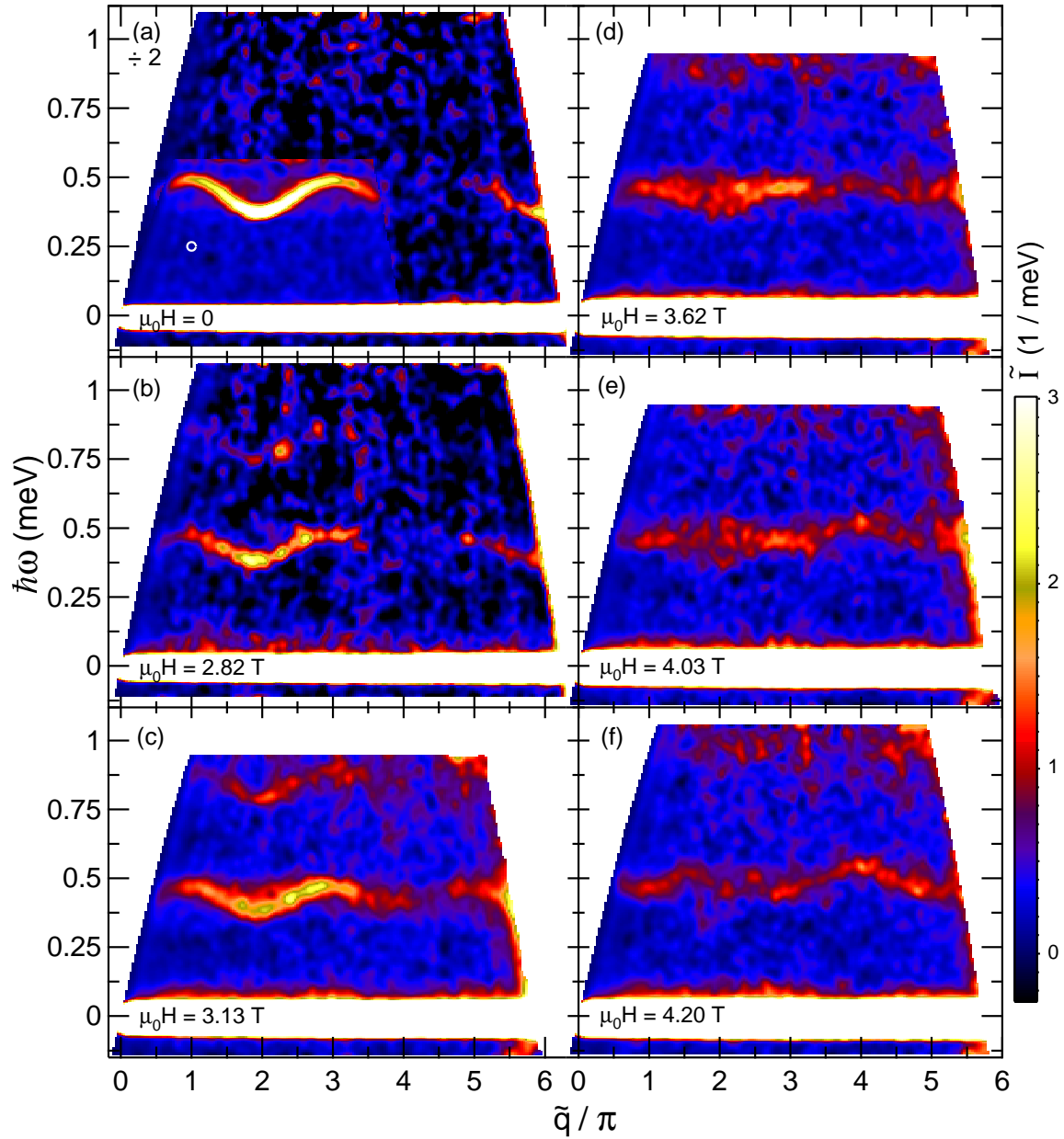}
\caption{\label{fig:cndcscontourdata}
(color online) Normalized scattering intensity, $\tilde{I}(\tilde{q}, \hbar\omega)$ for CN at $T=0.224$ K for (a) $H=0$, (b) $\mu_0 H=2.82$ T, (c) $\mu_0 H=3.13$ T, (d) $\mu_0 H=3.62$ T, (e) $\mu_0 H=4.03$ T, and (f) $\mu_0 H=4.2$ T from time-of-flight measurements.  The data in (a) are shown at half the intensity scale.  The  trapezoidal area in  (a) was covered by the high resolution Config. 3 while the surrounding area was measured with Config. 1. Data in (b) were measured with Config. 1. Data in (c), (d), and (e) were measured with Config. 2. Data in (f) were measured with Config. 4.  False color images of the scattering intensity were produced by binning the data onto an orthogonal grid in $\hbar\omega$ and $\tilde{q}$.  The data were then convolved with a normalized Gaussian with FWHM principal axes of $\hbar\omega=31$ $\mu$eV and $\tilde{q}/\pi = 0.125$. The corresponding FWHM ellipsoid is shown in (a), and approximates the average resolution ellipsoid.}
\end{figure*}

Excitations in the intermediate field range, $H_{c1} \leq H \leq H_{c2}$, were examined with the higher resolution time-of-flight instrument. Figure~\ref{fig:cndcscontourdata} shows data for magnetic fields $\mu_0 H=0$, $2.82$, $3.13$, $3.62$, $4.03$, and $4.2$ T.  The data were collected at  $T=0.224$~K, which exceeds the maximum N\'{e}el temperature for CN. The particular magnetic fields chosen correspond to rational fractions of the low-temperature saturation magnetization, $M_{\mathrm{sat}}$, in CN: at $\mu_0 H=\mu_0 H_{c1}=2.82$ T, $M=(0_+\epsilon) M_{\mathrm{sat}}$; at $\mu_0 H=3.13$ T, $M=\frac{1}{4}M_{\mathrm{sat}}$; at $\mu_0 H=3.62$ T, $M=\frac{1}{2}M_{\mathrm{sat}}$; at $\mu_0 H=4.03$ T, $M=\frac{3}{4}M_{\mathrm{sat}}$; and at $\mu_0 H=4.2$ T, $M=M_{\mathrm{sat}}$.\cite{hammarreichunpub}

We first compare the zero-field data (Fig.~\ref{fig:cndcscontourdata}(a)) to the SPINS data. The modes are sharper due to the enhanced energy resolution. This, as well as the lower temperature\cite{gxu00}  leads to the increased peak scattering intensity.  For the Shastry-Sutherland spin dimer system $\rm SrCu_2(BO_3)_2$ the singlet triplet transition near 3 meV is split by 0.4 meV or  13\%. For comparison Fig.~\ref{fig:cndcscontourdata}(a) places an upper limit of 0.02 meV or 5\% on any zero-field peak splitting. This indicates the Heisenberg approximation to intra-dimer spin interactions is excellent for CN.

At $\mu_0 H=\mu_0 H_{c1}=2.82$ T, Fig.~\ref{fig:cndcscontourdata}(b), the three modes of the zero field triplet are split.  The central, $S_{z}=0$, mode is at the same location as in zero-field.  The $S_{z}=\pm1 $ modes are less intense and Zeeman split to higher and lower values of $\hbar\omega$.  The $S_{z}=-1$ mode lies at $\hbar\omega \approx 0.75$ meV, while the $S_{z}=1$ mode is manifest in increased scattering intensity near the incoherent elastic line.

At higher fields, the upper mode continues to increase in energy while the central energy mode at $\hbar\omega \approx 0.45$ meV changes its $\tilde{q}$-dependence and the intensity distribution moves to slightly higher values of $\hbar\omega$.  At $\mu_0 H=\mu_0 H_{c2}=4.2$ T, Fig.~\ref{fig:cndcscontourdata}(f), the $\tilde{q}$ dependence of both the peak position and intensity associated with the central mode -- considered as a function of $\tilde{q}$ -- have changed phase.

  \begin{figure}
 \centering\includegraphics[scale=0.825]{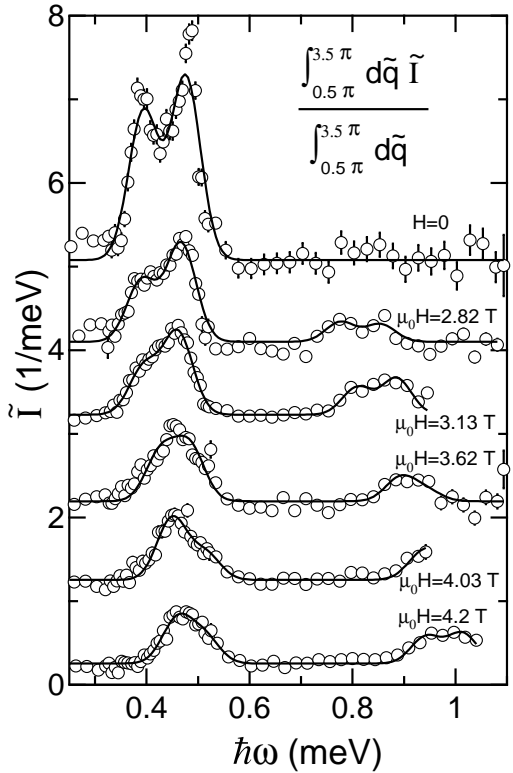}
 \caption{\label{fig:dcsintegratedintenvshw}
 Wave-vector averaged magnetic scattering intensity for  CN at $T=0.224$ K as a function  of $\hbar\omega$ and applied magnetic field from time-of-flight measurements.  The average covered the range from $\tilde{q}=\frac{1}{2}\pi$ to $\tilde{q} = \frac{7}{2}\pi$.  Data for different fields are offset vertically for presentation purposes.  Solid lines are the result of gaussian fits as described in the text. The $H = 0$ and $\mu_0 H = 2.82$ T data were combined from measurements made in Configs. 1 and 3. The $\mu_0 H = 3.13$ T and  $\mu_0 H = 4.03$ T data were acquired using Config 2. The $\mu_0 H = 3.62$ T data were combined from measurements made in Configs. 1, 2, and 3. The $\mu_0 H = 4.2$ T data are from measurements made in Config. 4.}
 \end{figure}

For an overview of the magnetic density of states versus field, the wave-vector integrated intensity ($\frac{1}{2}\pi<\tilde{q}< \frac{7}{2}\pi$) is shown in Fig.~\ref{fig:dcsintegratedintenvshw}.  In zero field, there are peaks associated with the Van-Hove singularities at the top and bottom of the magnetic excitation spectrum.  At the lower critical field, $\mu_0 H_{c1}=2.82$ T, these singularities become less pronounced and the upper $S_z=-1$ band appears as a rounded double peak at $\hbar\omega \approx 0.8$ meV.  With increasing field both bands shift to higher energies, with a greater rate for the higher energy $S_z=-1$ band.

 \begin{figure}
  \centering\includegraphics[scale=0.825]{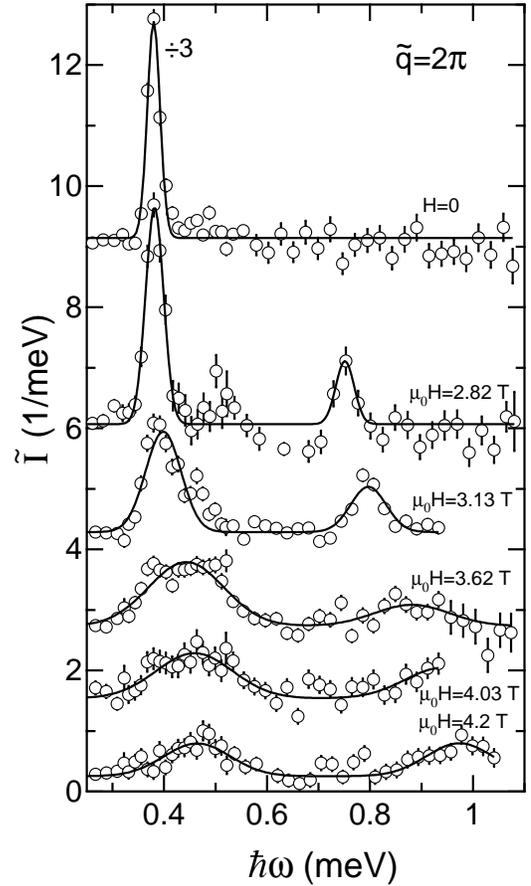}
 \caption{\label{fig:dcsivshbaromega2pi}
Constant $\tilde{q}=2\pi$ scattering intensity ($\delta\tilde{q}=0.13\pi$) as a function of $\hbar\omega$ and applied magnetic field for CN at $T=0.224$ K from time-of-flight measurements.  The zero-field data has been reduced in intensity by a factor of three for presentation purposes.  Solid lines are fits to Gaussians as described in the text.  Results have been offset vertically for presentation.  The $H = 0$ and $\mu_0 H = 2.82$ T data were combined from measurements made in Configs. 1 and 3. The $\mu_0 H = 3.13$ T and $\mu_0 H = 4.03$ T data are from Config 2. The $\mu_0 H = 3.62$ T data were
combined from measurements made in Configs. 1, 2, and 3. The $\mu_0 H = 4.2$ T data are from measurements made in Config. 4.}
\end{figure}

\begin{figure}
 \centering\includegraphics[scale=0.825]{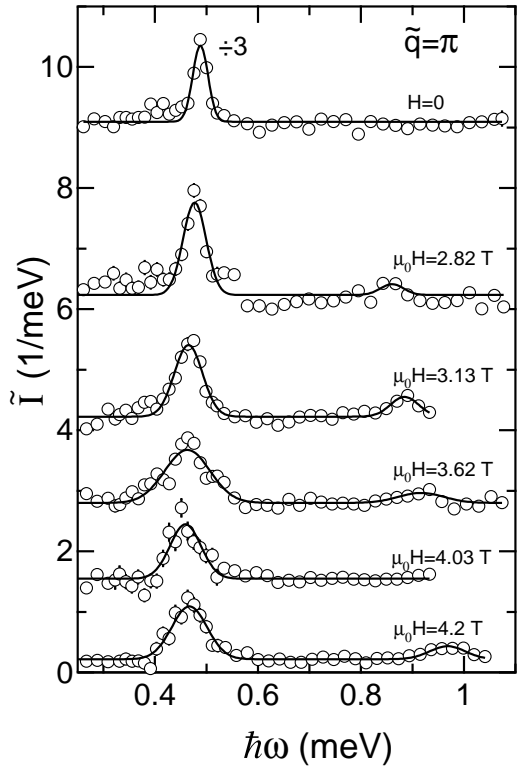}
\caption{\label{fig:dcsivshbaromegapi}
Constant $\tilde{q}=\pi$ scattering intensity ($\delta\tilde{q}=0.13\pi$) as a function of $\hbar\omega$ and applied magnetic field for CN at $T=0.224$ K from time-of-flight measurements.  The zero-field data has been reduced in intensity by a factor of three.  Solid lines are fits to Gaussians as described in the text.  Results have been offset vertically for clarity. The $H = 0$ and $\mu_0 H = 2.82$ T data were combined from measurements made in Configs. 1 and 3. The $\mu_0 H = 3.13$ T and $\mu_0 H = 4.03$ T data are from Config 2. The $\mu_0 H = 3.62$ T data were
combined from measurements made in Configs. 1, 2, and 3. The $\mu_0 H = 4.2$ T data are from measurements made in Config. 4.}
\end{figure}

Figures~\ref{fig:dcsivshbaromega2pi} and~\ref{fig:dcsivshbaromegapi} show constant $\tilde{q}=2\pi$ and $\tilde{q}=\pi$ scans extracted from the data shown in Fig.~\ref{fig:cndcscontourdata}.  In zero field at $\tilde{q}=2\pi$, there is a single peak at $\hbar\omega \approx 0.4$ meV corresponding to crossing the bottom of the magnetic band.  As the magnetic field is increased, this peak broadens, loses intensity, and moves to higher values of energy transfer.  An analogous evolution with field is observed for the upper branch of the Zeeman split triplet, which broadens and shifts from $\hbar\omega = 0.75$~meV to $\hbar\omega = 0.95$~meV upon raising the field from the lower critical field $\mu_0 H=\mu_0 H_{c1}=2.8$~T to the upper critical field $\mu_0 H=\mu_0 H_{c2}=4.2$~T.

For $\tilde{q}=\pi$ (Fig.~\ref{fig:dcsivshbaromegapi}) the single peak at $\hbar\omega \approx 0.45$ meV corresponds to crossing the top of the magnetic band in zero field.  With increasing field, this peak broadens, and also shifts slightly to lower energy transfer.  Just as for $\tilde{q}=2\pi$, the higher energy $S_z=-1$ branch becomes visible for $H>H_{c1}$ and shifts to higher $\hbar\omega$ with increasing fields.

\section{Analysis and Discussion}
\label{discussion}

While the resolution employed here is adequate to resolve the dispersion of gapped excitations, quasi-elastic features in the intermediate field regime are not well separated from the elastic nuclear scattering.\cite{stonethesis} Our discussion therefore focuses on gapped excitations, first covering fields below $H_{c1}$ and above $H_{c2}$, then the intermediate field regime. For context we start by presenting the results of exact diagonalization calculations for finite length alternating spin chains in an applied magnetic field.

\subsection{Exact Diagonalization}

 \begin{figure*}
 \centering\includegraphics[scale=0.8]{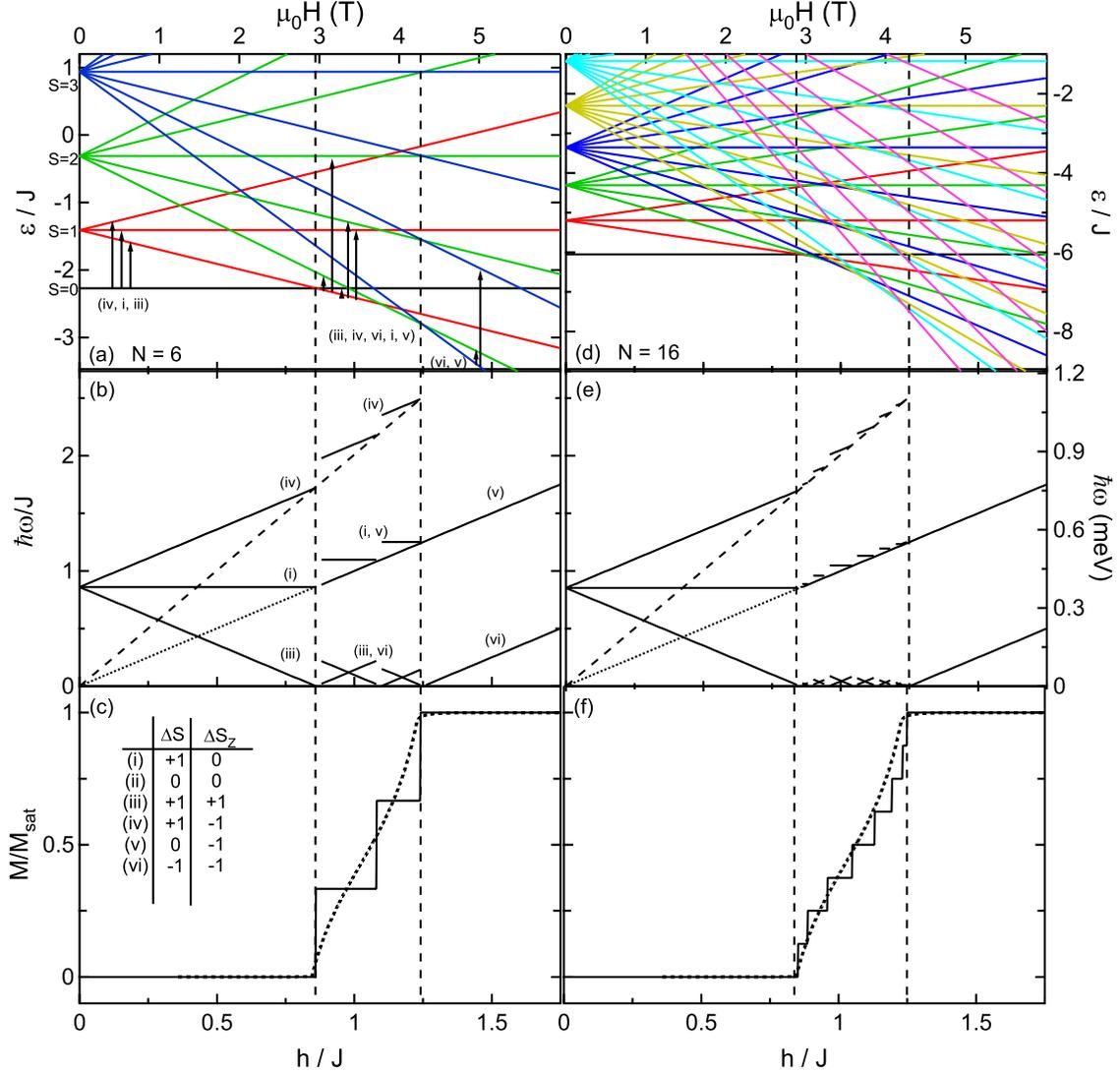}
 \caption{\label{fig:statediagram}
 (color online)  Zero temperature exact diagonalization calculations for $N=6$ (a-c) and $N=16$ (d-f) $\alpha = 0.25$ alternating antiferromagnetic Heisenberg spin chains as a function of applied magnetic field at zero wave-vector.  Panels (a) and (d) are the calculated field-dependent energy levels.  States illustrated in the same color belong to the same total spin multiplet.  Panels (b) and (e) are the lowest energy dipole  active transitions.  Dotted and dashed lines are extrapolations of the field dependence of excitations in the intermediate field regime.  Panels (c) and (f) are the calculated magnetization based on the total moment of the ground state.  The dotted line in these panels is the measured $T=0.1$ K field dependent magnetization of CN.\cite{hammarreichunpub}  Panels (a) and (b) also identify transitions in the basis of total spin $S$ and the $z$-component of spin $S_z$ using lower case Roman numerals (i) through (vi) as listed in the table in panel (c) and described in Ref.~\onlinecite{muller81}.  Reduced units are used for the bottom axes, and units of applied magnetic field for the case of $J_1 \approx 0.44$ are used for the top axes.}
 \end{figure*}

Exact diagonalization of finite length alternating spin-1/2 chains is instructive to understand the magnetic field dependent phase diagram and excitation spectra of CN. We performed such calculations for $N=6$ and $N=16$ member $S=\frac{1}{2}$ antiferromagnetic $\alpha=\frac{1}{4}$ alternating chains as a function of applied magnetic field at zero temperature.  The calculations yield the field-dependent evolution of the energy levels, the corresponding excitation spectra, and the ground state magnetization (Fig.~\ref{fig:statediagram}).

In zero field the Hamiltonian is rotationally invariant whereas in finite field it is invariant under rotations about the field axis. Correspondingly we label the dipole-active transitions from (i) to (vi) based on their respective total spin $S$ quantum number (appropriate for zero field) and the $z$-component of spin $S_z$ quantum numbers (appropriate in zero and finite fields).

The excitations in zero field are between a non-magnetic singlet ground state and a magnetic triplet ($S=1$).  A series of higher energy multiplets play a role in the intermediate field regime.  On application of an external magnetic field, the degeneracy of the three excitations for $H<H_{c1}$ shown in Fig.~\ref{fig:statediagram}(a), (b), (d) and (e) is lifted through Zeeman splitting of the total spin multiplets.  For larger magnetic fields, there exists a quantum critical point (QCP), $H_{c1}$, where the spin-gap is closed and magnetization develops.  In the intermediate field regime, a total of five non-trivial transitions are allowed as shown in Fig.~\ref{fig:statediagram}(a) and (b).

Upon increasing the chain length, as shown in the right panels of Fig.~\ref{fig:statediagram}, the field-dependence of the massive excitations in this regime becomes linear with an intercept at the origin.  Within the gapless portion of the phase diagram, inter-chain interactions are expected to produce a N\'eel-like ground state.  At higher fields there is a second QCP, $H_{c2}$, beyond which the system is fully polarized in a forced ferromagnetic state with two transverse spin-wave modes.  The calculated magnetization in these three field ranges agrees quite well with the measured low-temperature magnetization for CN (Fig.~\ref{fig:statediagram}(f)). In the following we shall discuss the experimental observations  in light of these exact diagonalization results.

\subsection{{\protect $H < H_{c1}$} and {\protect $H > H_{c2}$}}
The intensity modulation of the modes in Fig~\ref{fig:cncontourplotspins}(a)-(c) is not commensurate with the crystal lattice. This can be understood by examining the dynamic spin correlation function for a dimer antiferromagnet in conjunction with the SMA in the absence of anisotropy.\cite{hohenberg74,Ma92}  The SMA has been used with success to describe the zero field modes in copper nitrate such that the dynamic spin correlation function becomes\cite{gxu00}
\begin{equation}
{\mathcal S}({\bf Q},\hbar\omega) = \frac{J_1\langle {\bf S}_0\cdot{\bf S}_{{\bf d}_1} \rangle}{3\epsilon({\bf Q})}[1-\cos({\bf Q}\cdot{\bf d}_{1})]\nonumber\times\delta(\hbar\omega-\epsilon({\bf Q})),
\label{eq:sma}
\end{equation}
where ${\bf d}_{1}$ is the intra-dimer vector.  This expression describes the scattering associated with singlet to triplet transitions. In the magnetized state however, transitions can occur between states that originate from the zero field triplet. For an isolated spin pair these carry a structure factor of $\propto (1+cos(\bf Q\cdot \bf d_1))$.

For the case of a general spin cluster, the $[1-\cos({\bf Q}\cdot{\bf d}_{1})]$ structure factor in Eq.~\ref{eq:sma} needs to be modified to account for the initial $S$ and final $S^{\prime}$ spin quantum numbers of the excitation.\cite{furrer79,leuenberger86}  Based upon the spin states involved, we propose the phenomenological expression for the field dependent correlation function,
\begin{eqnarray}
{\mathcal S}({\bf Q},\hbar\omega)& = & \frac{J_1\langle {\bf S}_0\cdot{\bf S}_{{\bf d}_1} \rangle}{3\epsilon({\bf Q})}[1+(-1)^{S-S^{\prime}}\cos({\bf Q}\cdot{\bf d}_{1})]\nonumber\\
&&\times\delta(\hbar\omega-\epsilon({\bf Q})).
\label{eq:rpasqw}
\end{eqnarray}
Here $S$ and $S^\prime$ refer to the initial and final spin state of the transition labeled in accordance with the non-interacting limit. Equation~\ref{eq:rpasqw} retains the oscillator strength for each mode while introducing the anticipated dispersion from perturbation theory.  The expression is appropriate in the limit of $\alpha=J_2/J_1\rightarrow 0$ and for the forced ferromagnetic state as described shortly.

For $H < H_{c1}$, the low energy states of the antiferromagnetic alternating chain are the $S=0$ singlet ground state and the $S=1$ triplet excited state (Fig.~\ref{fig:statediagram}). The cosine term in Eq.~\ref{eq:rpasqw} has a negative prefactor for these transitions.  Figure~\ref{fig:cncontourplotspins}(d) is a fit to the zero field data based on equations \ref{eq:zeemandisp} through \ref{eq:rpasqw}.  The incoherent elastic background was modeled as a Gaussian and a Lorentzian peak as a function of energy transfer and a Lorentzian peak as a function of $\tilde{q}$ to account for an increase in background at smaller values of scattering angle.  This model spectrum has three fitting parameters: two exchange constants, $J_1$ and $J_2$, and an overall multiplicative prefactor.  The model accounts well for the scattering intensity and dispersion with $J_{1}=0.44(1)$ meV and $J_{2} =0.09(1)$ meV.  The values of intra- and interdimer exchange are consistent with the values obtained from measurements at similar temperatures described earlier and with the temperature dependent renormalization of the bandwidth determined by Xu \textit{et al}.\cite{gxu00}

To describe the data in Fig.~\ref{fig:cncontourplotspins}(b) we employed the Zeeman split dispersion as described in Eq.~\ref{eq:zeemandisp} with $\epsilon_{0}(\mathbf{Q})$ given by Eq.~\ref{eq:oneddisph0l} along with Eqs.~\ref{eq:rpasqw} and ~\ref{eq:intneut}.  Considering the potential for an anisotropic $g-$factor, prefactors for the $S_{z} \neq 0$ components were allowed to vary and this resulted in the excellent fit shown in Fig.~\ref{fig:cncontourplotspins}(e). The corresponding values of $J_{1}=0.43(1)$~meV, $J_{2}= 0.071(7)$~meV are consistent with the zero field fits. The intensitiies of the lower and upper modes respectively were found to be 39(1)$\%$ and 47(2)$\%$ of the $S_{z}=0$ corresponding on average to $\delta g=0.17(1)$ , which is consistent with $g=\sqrt{g_{b}^{2}+g_{\perp}^{2}}=2.22$ derived from bulk measurements.\cite{bonnerprb1983}

For $H>H_{c2}$, the $\tilde{q}$ dependence of the $\omega$-integrated intensity for  the lower and upper modes can be understood in terms of  Eqs.~\ref{eq:rpasqw} and \ref{eq:intneut} in conjunction with the transitions illustrated in Fig.~\ref{fig:statediagram}.  From our finite chain calculations, the change in total spin from the ground state to the lowest excited state for the ferromagnetic phase is odd, $\Delta S=-1$, while the change in total spin to the next highest excited state is even, $\Delta S = 0$.  This leads to the phase difference in the intensities of the two modes in the $\mu_0 H = 7$ T spectrum through the $(-1)^{S-S^{\prime}}$ term in Eq.~\ref{eq:rpasqw}.

For a simulation of the high field data we use the Holstein-Primakoff spinwave calculation, to calculate the dispersion of a 1d alternating ferromagnet as
\begin{eqnarray}
\epsilon_{FSW}(\mathbf{Q}) &=&  g \mu_{B} (H-H_{c2}) + \frac{1}{2}(J_{1}+J_2)\nonumber \\
  & \pm &\frac{1}{2}\sqrt{J_{1}^{2}+J_{2}^{2}+2J_{1}J_{2}\cos({\bf Q\cdot u}))}.
\label{eq:fmdispcn}
\end{eqnarray}
where $S$ is the spin of the magnetic ions, $J_1$ and $J_{2}$ are the intra- and inter-dimer exchange and $\mathbf{u}$ is the inter-dimer vector.  Here the applied magnetic field term has been offset by the upper critical field.
The dispersion in Eq.~\ref{eq:fmdispcn} is identical to that found via a bosonization representation of the excitations of the alternating ferromagnetic chain.\cite{huang91}

The data in Fig.~\ref{fig:cncontourplotspins}(c) were fit using Eqs.~\ref{eq:rpasqw} -~\ref{eq:fmdispcn} and a comparison between model and data is shown in Fig.~\ref{fig:cncontourplotspins}(f). The best fit exchange parameters $J_{1}=0.46(2)$ meV, $J_2 = 0.07(1)$ meV are consistent with measurements from the singlet ground state. The fit included a multiplicative scale factor corresponding to the ratio of  intensities of the upper band to the lower band to account for the $g-$factor.

CN is thus well described by the alternating chain model for magnetic fields below the lower critical field and above the upper critical field.  The fitted exchange constants found for $H<H_{c1}$ and $H>H_{c2}$ are similar even though the non-negligible temperature of the measurements may affect the results near the critical fields.  The weighted average of the alternating chain exchange constants determined from high and low field measurements in the temperature range $1.2 \leq T \leq 1.8$ K are $\bar{J_1} = 0.4430(4)$ meV and $\bar{J_2} = 0.0849(7)$ meV. The $\tilde{q}$ dependence of the low energy modes found in the exact diagonalization calculations is consistent the measured spectra below $H_{c1}$ and above $H_{c2}$.

\begin{figure}
  \centering\includegraphics[scale=0.95]{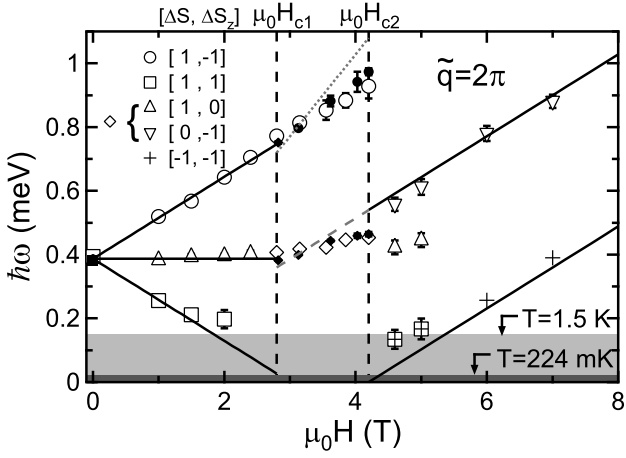}
 \caption{\label{fig:peaklocspinsdcs2pi}
 Fitted peak position in the $\tilde{q}=\pi$ magnetic neutron scattering spectrum of CN as a function of applied magnetic field  for $T=1.8$ K (open symbols, triple-axis  measurements) and $T=0.224$ K (closed symbols,  time-of-flight measurements). The raw data and fits are shown in Figs.~\ref{fig:qeq2pi_spins} and~\ref{fig:dcsivshbaromega2pi}.  $\mu_0 H_{c1}=2.8$ T and $\mu_0 H_{c2}=4.2$ T are indicated by dashed vertical lines.  For excitations near $\hbar\omega\approx 0.45$ meV at intermediate fields, fits were performed with an additional FWHM parameter $2\Gamma$ added in quadrature to the resolution width. The time-of-flight constant $\tilde{q}$ scans in the intermediate field range shown in Figs.~\ref{fig:dcsivshbaromega2pi} and ~\ref{fig:dcsivshbaromegapi} are fit to Gaussian peaks.  The light [dark] shaded area corresponds to $k_{B}T=0.155 [0.019]$ meV.  Symbol styles correspond to different excitations as noted in the legend.  Open square  with cross inset for $\mu_0 H=4.6$ and $5$ T corresponds to lower-energy excitation determined from higher energy thermally populated transitions as described in the text.  Solid lines are theoretical mode energies based on finite chain calculations.  Dotted and dashed  grey lines in the intermediate field regime are the limiting behavior of mode energies based on finite chain calculations.}
 \end{figure}

For an overview of the field dependence of the excitation spectrum we fit the constant $\tilde{q}=\pi$ and $\tilde{q}=2\pi$ of Figs.~\ref{fig:qeq2pi_spins} and ~\ref{fig:qeq2nplus1pi_spins} to sums of resolution limited Gaussians.  The $\hbar\omega$-dependent background  determined from analysis of the full data set of Fig.~\ref{fig:cncontourplotspins} was used. The field-dependent peak positions for $\tilde{q}=2\pi$  are shown in Fig.~\ref{fig:peaklocspinsdcs2pi}. We plot the field dependence of the calculated $H<H_{c1}$ and $H>H_{c2}$ excitations at $\tilde{q}=2\pi$ as solid lines in Fig.~\ref{fig:peaklocspinsdcs2pi} with $g=\sqrt{g_{b}^{2}+g_{\perp}^{2}}=2.22$ \cite{bonnerprb1983} using the low-temperature values of the exchange constants.\cite{gxu00} There is very good agreement between the measured and calculated field dependence.

While the modes observed for $\mu_0 H \geq 6$ T data are consistent with Eq.~\ref{eq:fmdispcn}, spectra for fields between $\mu_0 H_{c2}$ and $\mu_0 H = 5$ T are not.  There are two resolution limited modes (represented by inverted and upright triangles in Fig.~\ref{fig:peaklocspinsdcs2pi}), but they are not separated by the same energy as for the $\mu_0 H \geq 6$ T data.   Immediately above $\mu_0 H_{c2}$, The higher energy mode is consistent with theory, but the lower energy modes occur at elevated values of $\hbar\omega$.  This can be understood by considering the temperatures used in the measurements.  Just above $\mu_0 H_{c2}$,  low energy states are thermally populated due to the finite temperature of the measurement: $T=1.4$~K.  The additional excitation  observed for $\mu_0 H=4.6$ and $5$ T are due to scattering from the excited states rather than the ground state, and we label them as $\Delta S$=1, and $\Delta S_{z}$=0.  The energy of the zero temperature lowest energy excitation out of the ground state then corresponds to the difference in energies of the two modes observed.  These values are plotted as crosses inside of open squares in Fig.~\ref{fig:peaklocspinsdcs2pi}, and are consistent with the field dependent ferromagnetic dispersion above $H_{c2}$.

\begin{figure}
\centering\includegraphics[scale=0.9]{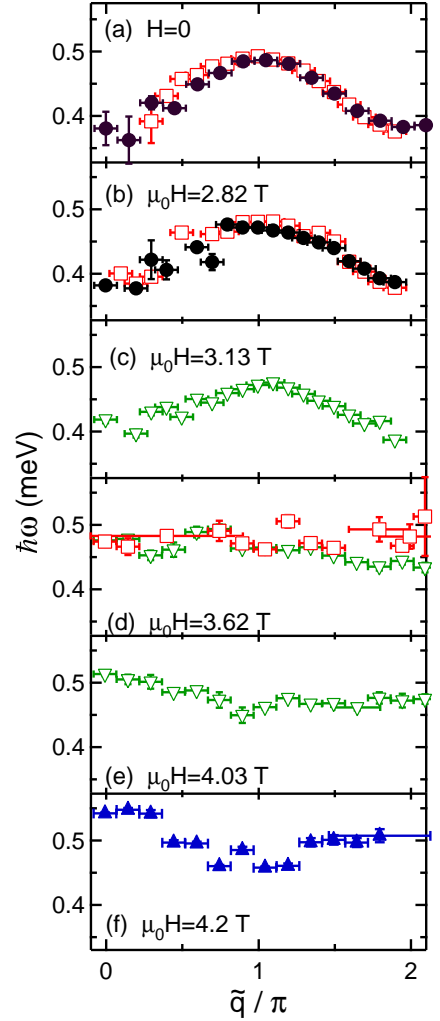}
\caption{\label{fig:cndcsdispvsfield}(color online)
Zero field (a) and intermediate field (b)-(f) dispersion relations for the $S_z=0$ mode near $J_1=0.44$~meV extracted by fitting the time-of-flight measurements shown in Fig.~\ref{fig:cndcscontourdata}(a)-(f) as described in the text.  The horizontal error bar represents the range of integration in $\frac{\tilde{q}}{\pi}$ used for the constant wave-vector scans which were fit to extract the dispersion values.  Multiple data sets for a given magnetic field were extracted from the different configurations used during the measurement.  The blue triangles correspond to fitting the data from configuration 4.  The green inverted triangles were measured in configuration 2. The solid(open) black(red) circles(squares) correspond to fits to data acquired in configuration 3(1)}
\end{figure}

\subsection{Intermediate fields, {\protect $H_{c1}  < H < H_{c2}$}}

While the magnetization is constant for fields below $H_{c1}$ and above $H_{c2}$ it varies continuously with field between these critical values. The implication is a gapless excitation spectrum that cannot support truly long-lived magnetic excitations. Near the elastic line a spectrum resembling that of the gapless spin-1/2 chain is anticipated but the instrumentation utilized here cannot resolve these features from the intense incoherent elastic nuclear scattering. Fig.~\ref{fig:cncontour3p55spins}~(a) however,  shows the band of magnetic excitations near $J_1$ is also qualitatively different in the intermediate field regime. While a sinusoidal dispersion remains, the corresponding amplitude is substantially reduced (Fig.~\ref{fig:cncontour3p55spins}~(b)). This can be seen as integral to the reversion of the dispersion relation from that of magnons within a singlet  for  $H<H_{c1}$ where $\tilde{q}=2n\pi$ is a minimum to magnons within a ferromagnet where $\tilde{q}=2n\pi$ is a maximum in the dispersion relation for this mode. This change in phase of the dispersion relation is clearly seen in Fig.~\ref{fig:cndcsdispvsfield} which summarizes the dispersion relation for all fields probed in this experiment through the positions of peaks in gaussian fits to constant-$\tilde{q}$ cuts through the data.  The constant-$\tilde{q}$ cuts were made after averaging about the $\tilde{q}=2\pi$ wave-vector to improve counting statistics.

More than a gradual change in the dispersion relation, Fig.~\ref{fig:cncontour3p55spins}(a), Fig.~\ref{fig:qeq2pi_spins}, and Fig.~\ref{fig:cndcscontourdata}(c)-(f) show the  character of the magnetic excitation spectrum is modified in the  intermediate field range. The spectrum is broadened beyond resolution and while thermal effects surely play a role for $T=1.8$~K the broadening clearly visible in Fig.~\ref{fig:cndcscontourdata}(d)-(e) for $T=0.224$~K is an intrinsic feature of the low $T$ spectrum.

For $H<H_{c1}$ the excitation near $J_1$ is associated with an $S_z=0$ magnon propagating through the singlet ground state. The absence of damping there indicates coherent propagation through the singlet which might be considered a consequence of the lack of lower energy states for the magnon to decay into. The substantial damping for $H_{c1}  < H < H_{c2}$ is evidence that the ground state has fundamentally changed character into one that offers a plethora of decay processes for the $S_z=0$ magnon. The incoherent magnon propagation is consistent with a Luttinger liquid ground state at intermediate fields. Such a state features a continuum of low energy two-spinon excitations that can exchange linear and angular momentum with the magnon and lead to the incoherent nature of the $S_z=0$ state that we observe.

\section{Conclusions}
\label{conclusions}
We have presented a comprehensive study of the wave-vector dependence of the gapped excitation spectrum of the 1d quantum spin liquid CN.  A careful examination of the three dimensional dispersion of the zero field singlet-triplet excitations has established that the alternating spin chains in this material extend along the $[111]$ or $[1\bar{1}1]$ real space direction and that the leading inter-chain interaction is $J^\prime_{2}=-6(1)~\mu$eV.

Below the lower critical field and above the upper critical field, the excitations are well described as being magnon-like wave packets traveling through respectively a singlet ground state and a forced ferromagnet. While the phase of the $\tilde{q}-$dependent scattering intensity and dispersion relation is reversed the data can be described by a single set of exchange constants, $\bar{J_1} = 0.4430(4)$ meV and $\bar{J_2} = 0.0849(7)$ meV. Exact diagonalization calculations for finite length alternating spin chains provide an excellent guide to understanding the field dependent data.

The intermediate field regime is considerably more complicated as it features a gapless excitation spectrum in the low $T$ limit. Our measurements of the $\hbar\omega\approx J_1$ mode show that the change in phase of its dispersion relation is accompanied by the loss of coherence for  $S_z=0$ magnon propagation. Employing the $S_z=0$ magnon to probe the low energy excitation spectrum, this result provides indirect evidence for a gapless continuum of excitations from the partially magnetized ground state. Given the good knowledge of the spin-hamiltonian for CN, it would be of great interest to see the numerical and analytical methods of one-dimensional quantum magnetism attempt to account for these observations.

In the mean time the next experimental step in the process of exploring magnetic exceptions of magnetized CN is a high resolution study of inelastic scattering up to approximately 0.25 meV  with a resolution better than $15$~$\mu$eV that is required to separate the magnetic scattering intensity from the intense incoherent elastic line of this coordination polymer quantum magnet.

\section{Acknowledgments}
A portion of this research at ORNL was sponsored by the Scientific User Facilities Division, Office of Basic Energy Sciences, U.S. Department of Energy.  CB was supported by the US Department of Energy, Office of Basic Energy Sciences, Division of Materials Sciences and Engineering under Award DE-FG02-08ER46544. This work utilized facilities supported in part by the National Science Foundation under Agreement No. DMR-0454672.


\begin{thebibliography}{99}
\bibitem{spacegroupnote}
The space group is incorrectly listed in reference~\onlinecite{morosin70}
as $I2/a$.
\end{thebibliography}
\end{document}